\documentclass[a4paper,fleqn,usenatbib]{mnras}

\usepackage{amsmath}	
\usepackage{amssymb}	

\usepackage{mathptmx}
\usepackage{txfonts}

\usepackage[T1]{fontenc}
\usepackage{ae,aecompl}

\usepackage{graphicx}	

\usepackage{bm}

\title[ZVI in protoplanetary discs]{On the survival of zombie vortices in protoplanetary discs}

\author[G. Lesur \& H. Latter]{
Geoffroy R. J. Lesur$^{1,2}$\thanks{E-mail: geoffroy.lesur@univ-grenoble-alpes.fr}
Henrik Latter,$^{3}$
\\
$^{1}$CNRS, IPAG, F-38000 Grenoble, France\\
$^{2}$Univ. Grenoble Alpes, IPAG, F-38000 Grenoble, France\\
$^{3}$DAMTP, University of Cambridge, CMS, Wilberforce Road, Cambridge CB3 0WA, UK
}

\date{Accepted XXX. Received YYY; in original form ZZZ}

\pubyear{2015}

\begin{document}
\label{firstpage}
\pagerange{\pageref{firstpage}--\pageref{lastpage}}
\maketitle

\begin{abstract}
Recently it has been proposed that the zombie vortex instability (ZVI)
could precipitate hydrodynamical activity and angular momentum transport
in unmagnetised regions of protoplanetary discs, also known as ``dead
zones''. 
In this letter we scrutinise, 
with high resolution 3D spectral simulations,
the onset and survival of this
instability in the presence of viscous and thermal physics. 
First, we find that the ZVI is strongly dependent on the nature of the
viscous operator. Although the ZVI is easily obtained with
hyper-diffusion, it is difficult to sustain with physical
(second order) diffusion operators up to Reynolds numbers as high as
$10^7$. This sensitivity is probably due to the ZVI's reliance on
critical layers, 
whose characteristic lengthscale, structure, and dynamics are controlled
by viscous diffusion.
Second, we observe that the ZVI is sensitive to radiative processes,
and indeed only operates when the Peclet number is greater than a
critical value $\sim 10^4$, or when the cooling time is longer than $\sim 10\Omega^{-1}$.
As a consequence, the ZVI struggles to appear at $R\gtrsim 0.3\,AU$ in
standard $0.01\,M_\odot$ T Tauri disc models, though
younger more massive disks provide a more hospitable environment.
Together these results question the prevalence
of the ZVI in protoplanetary discs.

\end{abstract}




\section{Introduction}

The origin of angular momentum transport in accretion discs, 
especially protoplanetary discs, is a long-standing issue in the
astrophysical community. Angular momentum transport is the mechanism
that governs the global dynamics of the gas, in particular its
accretion onto the central star. It is therefore especially important
 if one is to predict the long-term evolution and structure of protoplanetary discs.

The magnetorotational instability (MRI, \citealt{BH91}) is believed to
be the main driver of angular momentum transport in accretion
discs. By sustaining three-dimensional MHD turbulence, the MRI
transports angular momentum outwards and leads to mass accretion at
rates compatible with observations. It is far from assured, however, that
cold protoplanetary discs are sufficiently ionised to sustain MHD
turbulence. This has led to the concept of ``dead
zones'' \citep{G96},
internal regions of the disk where the MRI is quenched. 
The question of angular momentum
transport in dead zones is highly debated, and 
in which hydrodynamical instabilities are likely to be key \citep{T14}. 

The radial Keplerian rotation profile of astrophysical discs is known
to be hydrodynamically stable, both linearly and non-linearly
\citep{L05,EJ14}. However, additional physics, such as cooling,
heating, and
stratification, could unleash new hydrodynamical instabilities. 
In recent years, several have been
identified, including the subcricital
baroclinic instability (SBI, \citealt{P07,LP10}), the vertical shear
instability (VSI, \citealt{NG13}), the convective overstability
\citep{KH14} and more recently the zombie vortex instability (ZVI) 
 which appears in rotating shear flows exhibiting a stable vertical stratification.

The ZVI was first observed (but not clearly identified as such) in the
anelastic simulations of \citealt{BM05} and was subsequently isolated
by \citealt{MP13} using Boussinesq spectral simulations. This
instability, of nonlinear nature, produces ``self-replicating''
vortices thanks to the
excitation of very thin critical layers. The ZVI also
appears in compressible simulations with various initial
conditions. It has been proposed, but not yet demonstrated, that the
excitation of spiral density waves by zombie vortices could lead to
significant angular momentum transport in dead zones \citep{MP15},
thereby solving the angular momentum transport problem in these
regions. However, the physical mechanism driving the instability remains
mysterious. The existence and excitation of critical layers by a
perturbation is a well known linear mechanism in shear flows
\citep{DR81}, but their non-linear saturation and 
spontaneous transformation into new vortices is largely unexplained.
Because diffusive
physics determines the layer's structure and evolution, 
the nature of viscosity (and whether it
is physical or numerical) should be a fundamental ingredient in any theory.

In this paper, we scrutinise the foundations of the ZVI with our focus
squarely on the role of viscosity and cooling. We first
present our physical model and numerical methods, which are very
similar to \citealt{MP13}. We then move to the question of the
physical convergence of the ZVI as a function of the viscous
operator. We also explore the dependence of the ZVI on cooling, and
compare our results to realistic protoplanetary disc models. 
 We finally summarise our results and propose future routes of research into the ZVI.

\section{Methods}
\subsection{Physical Model}

We represent the local dynamics of the disc with the shearing box approximation. To further simplify the dynamics, 
we employ incompressibility but include vertical buoyancy effects via the Boussinesq approximation. In this framework, the equations of motion read
\begin{align}
\nonumber \frac{\partial \bm{u}}{\partial t}+\bm{u\cdot\nabla u}&=-\bm{\nabla}\Pi-2\bm{\Omega\times u}+2\Omega S x\bm{e}_x\\
&\qquad -N^2\theta\bm{e}_z+\nu\bm{\Delta u}+(\nu_6 \bm{\nabla}^2)^3\bm{u}, \\
\frac{\partial \theta}{\partial t}+\bm{u\cdot\nabla }\theta&=u_z+\chi \Delta \theta+(\chi_6\bm{\nabla}^2)^3\theta-\frac{\theta}{t_c},\\
\label{eq:divv}\bm{\nabla \cdot u}&=0.
\end{align}
In the above formulation we have defined the local rotation frequency $\Omega$, the shear rate $S$, the Brunt-Vaissala frequency $N$ and the generalised pressure $\Pi$, which allows us to satisfy the incompressible condition (\ref{eq:divv}). In addition, we have introduced several explicit diffusion operators: the usual second order viscosity $\nu$ and thermal diffusivity $\chi$ are supplemented with $6^\mathrm{th}$ order "hyper-diffusion" operators with coefficients $\nu_6$ and $\chi_6$. Hyper-diffusion has no real physical motivation but can be useful numerically to reduce diffusion on large scales without accumulating energy at the grid scale. Note that a similar hyper-diffusion operator was used by \citealt{MP15}. Finally, we have added a newtonian cooling in the form of a constant thermal relaxation time $t_c$.

The above set of equations admits a simple solution of pure shear flow $\bm{u}_0=-Sx\bm{e}_y$. In the following, we define perturbations (not necessarily small) to this global shear flow $\bm{v}=\bm{u}-\bm{u}_0$. 

The equations of motions are supplemented by a set of periodic boundary conditions in the $y$ and $z$ directions. In the $x$ direction, we use shear-periodic boundary conditions, following \citealt{HGB95}.

\subsection{Dimensionless numbers and units}

The set of equations above includes several dynamical timescales that
can be usefully compared via appropriate dimensionless numbers. We define and use the following ones:
\begin{itemize}
\item the Rossby number $q=S/\Omega$. In Keplerian accretion discs $q=3/2$, which we will be the framework of this letter.
\item the Froude number $\mathrm{Fr}=N/\Omega$. In this work, we will always assume $\mathrm{Fr}=2$ which corresponds to the fiducial case studied by \citealt{MP13} of a moderately stratified flow. Note however that expected Froude numbers in protoplanetary discs are somewhat lower than this value, with $\mathrm{Fr}\simeq 0.3$ \citep{DM05}. Our setup therefore represents an upper bound on the amplitude of stratification effects.
\item the Reynolds number $\mathrm{Re}=\Omega L^2/\nu$ compares the amplitude of nonlinear advection terms to viscous diffusion. Equivalently, we define a Reynolds number based on hyperviscosity $\mathrm{Re}_6=\Omega^{1/3} L^2/\nu_6$
\item the Peclet number $\mathrm{Pe}=\Omega L^2/\chi$ compares nonlinear advection to thermal diffusion. As for the Reynolds number, we also define a hyperdiffusion Peclet number $\mathrm{Pe}_6$.
\item the dimenionsless cooling time $\tau=t_c\Omega$.
\end{itemize} 

Unless mentioned otherwise, we use $\Omega^{-1}$ as our time unit and the box size $L$ as our length unit. 

\subsection{Numerical technique}

We employ Snoopy to integrate the equations of motion. Snoopy is a spectral code using a Fourier decomposition of the flow to compute spatial derivatives. Time integration is performed using a low storage 3rd order Runge-Kutta scheme. Diffusive operators are solved by an implicit operator which maintains the 3rd order accuracy of the scheme. To avoid spectral aliasing due to the quadratic nonlinearities, we use a standard 2/3 anti-aliasing rule when computing each nonlinear term. The code and ZVI setup is freely available on the author's website.

In this letter, we use two sets of initial condition: single vortex initial conditions (runs labels ending with ``v'') and Kolmogorov-like noise (runs labels ending with ``k''). 

Our single vortex initial condition is similar to \citealt{MP13} with an isolated gaussian vortex centred at the origin of the box with a size $\sigma$ and a velocity amplitude $v_0$. The initial perturbation reads
\begin{align}
\nonumber v_x^0(\bm{x})&=y v_0/\sigma\exp\Big[-(x^2+y^2+z^2)/\sigma^2\Big],\\
\nonumber v_y^0(\bm{x})&=-x v_0/\sigma\exp\Big[-(x^2+y^2+z^2)/\sigma^2\Big].
\end{align} Our simulations start with $\sigma=0.07$ and $v_0=0.03$ in order to get results close to \citealt{MP13}. This perturbation corresponds to a stratified anticyclonic vortex with a vertical vorticity $\omega_z=\partial_x v_y-\partial_y v_x\simeq -0.8$. 

When using Kolmogorov-like noise, we randomly excite each velocity wavenumber isotropically in phase and amplitude and set the energy spectrum to $E(k)\propto k^{-5/3}$. We normalise our initial conditions so that $\sqrt{\langle \bm{v}^2\rangle}=4\times 10^{-2}$ at $t=0$.

\section{Results}
\subsection{Fiducial case and and hyperdiffusion}

\begin{table}
\centering
\setlength{\tabcolsep}{4pt}
\begin{tabular}{ l ccccccc}
\hline
Run & $\mathrm{Re}$ & $\mathrm{Pe}$ & $\mathrm{Re}_6$ & $\mathrm{Pe}_6$ & $\tau $ &Resolution & ZVI\\
\hline
\hline
h-256-v	& $\infty$ & $\infty$ & $5\times 10^5$	& $5\times 10^5$ & $\infty$ &$256^3$ & yes\\
h-1024-v	& $\infty$ & $\infty$ & $5\times 10^5$	& $5\times 10^5$ & $\infty$&$1024^3$ & yes\\
v6-1024-v	& $10^6$ 	& $10^6$ & $\infty$			& $\infty$ 		& $\infty$&$1024^3$ & no\\
v7-1024-v	& $10^7$ 	& $10^7$ & $\infty$			& $\infty$ 		& $\infty$&$1024^3$ & no\\
\hline
d5-256-v	& $\infty$ 	& $6.4\times 10^5$ & $5\times 10^5$	& $\infty$ 		& $\infty$& $256^3$ & yes\\
d4-256-v	& $\infty$ 	& $1.0\times 10^5$ & $5\times 10^5$	& $\infty$ 		& $\infty$& $256^3$ & yes\\
d4-256-k	& $\infty$ 	& $8.0\times 10^4$ & $5\times 10^5$	& $\infty$ 		& $\infty$& $256^3$ & yes\\
d3-256-v	& $\infty$ 	& $5.0\times 10^4$ & $5\times 10^5$	& $\infty$ 		& $\infty$& $256^3$ & yes\\
d3-256-k	& $\infty$ 	& $4.0\times 10^4$ & $5\times 10^5$	& $\infty$ 		& $\infty$& $256^3$ & yes\\
d2-256-v	& $\infty$ 	& $2.0\times 10^4$ & $5\times 10^5$	& $\infty$ 		& $\infty$& $256^3$ & no\\
d2-256-k	& $\infty$ 	& $2.0\times 10^4$ & $5\times 10^5$	& $\infty$ 		& $\infty$& $256^3$ & no\\
d1-256-v	& $\infty$ 	& $1.0\times 10^4$ & $5\times 10^5$	& $\infty$ 		& $\infty$& $256^3$ & no\\
d1-256-k	& $\infty$ 	& $1.0\times 10^4$ & $5\times 10^5$	& $\infty$ 		& $\infty$& $256^3$ & no\\
\hline
t5-256-v	& $\infty$ 	& $\infty$ & $5\times 10^5$	& $\infty$ 		& 128 & $256^3$ & yes\\
t4-256-v	& $\infty$ 	& $\infty$ & $5\times 10^5$	& $\infty$ 		& 64 & $256^3$ & yes\\
t4-256-k	& $\infty$ 	& $\infty$ & $5\times 10^5$	& $\infty$ 		& 64 & $256^3$ & yes\\
t3-256-v	& $\infty$ 	& $\infty$ & $5\times 10^5$	& $\infty$ 		& 32 & $256^3$ & yes\\
t3-256-k	& $\infty$ 	& $\infty$ & $5\times 10^5$	& $\infty$ 		& 32 & $256^3$ & yes\\
t2-256-v	& $\infty$ 	& $\infty$ & $5\times 10^5$	& $\infty$ 		& 16 & $256^3$ & no\\
t2-256-k	& $\infty$ 	& $\infty$ & $5\times 10^5$	& $\infty$ 		& 16 & $256^3$ & no\\
\hline 
\end{tabular}

\vspace{1mm}
\caption{\label{tab:runs}List of simulations discussed in this letter.} 
\end{table}
\vspace{-0.2cm}

\begin{figure}
	\includegraphics[width=0.9\linewidth]{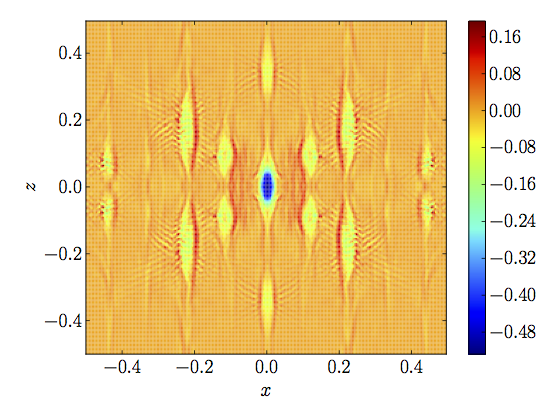}
	\vspace{-0.5cm}
    \caption{Vertical vorticity $\omega_z$ in a $x-z$ cut of our fiducial simulation with $\mathrm{Re}_6=\mathrm{Pe}_6=5\times 10^{5}$ at $t=500$. Similarly to \citealt{MP13}, we observe the formation and replication of anticyclonic vortices on a fixed lattice.}
    \label{fig:rep_vort}
\end{figure}

\begin{figure}
	\includegraphics[width=0.9\linewidth]{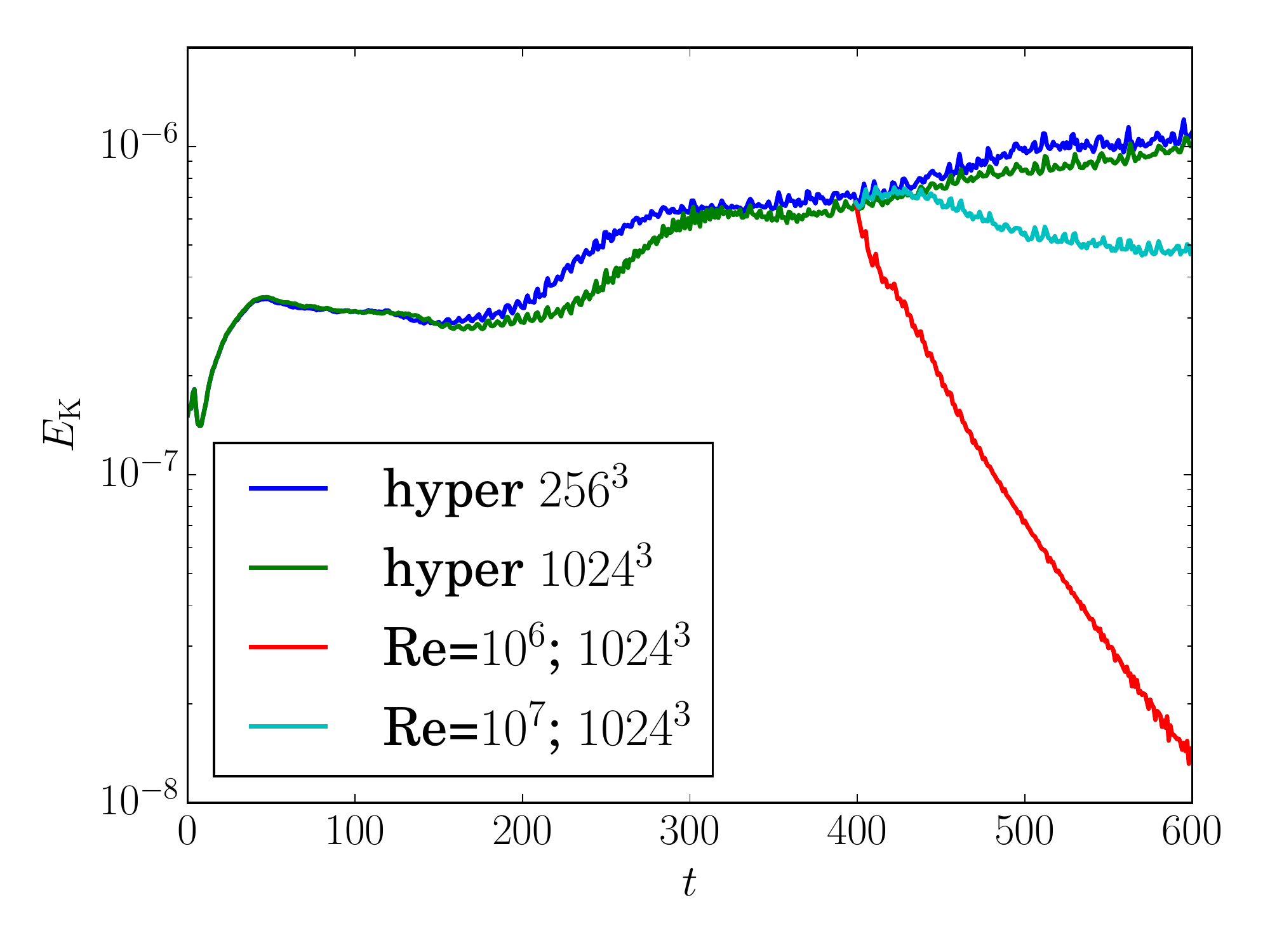}
	\vspace{-0.5cm}
    \caption{Volume averaged kinetic energy as a function of time for several simulations at $\mathrm{Fr}=2$ and $q=3/2$. The blue curve corresponds to our fiducial case. High resolution runs including diffusion are restarted from the hyper-diffusion run at $t=400$. }
    \label{fig:energy}
\end{figure}

We first introduce our fiducial model h-256 (see Tab.~\ref{tab:runs}) which essentially reproduces the results  of \citealt{MP13}. We choose a resolution of $256^3$ Fourier modes in a cubic box representing a Keplerian disc with $q=3/2$ and $\mathrm{Fr}=2$. No viscosity nor diffusion is imposed, $\nu=\chi=0$. We instead use 6th order hyper-diffusion to dissipate energy at small scales that would otherwise accumulate, spectral codes being inherently energy conserving schemes. We set $\mathrm{Re}_6=\mathrm{Pe}_6=5\times 10^{5}$. This simulation allows us to reproduce the main results of \citealt{MP13}: self replicating vortices on a fixed lattice (Fig.~\ref{fig:rep_vort}), and a growth in kinetic energy $E_K\equiv \langle \bm{v}^2/2\rangle $ associated with these vortices (Fig.~\ref{fig:energy}, blue line). As expected, new vortices appear at critical layers defined, from the initial vortex, by $x_c=\pm \mathrm{Fr}L /(2\pi m q)\simeq \pm 0.21/m$.

\begin{figure}
	\includegraphics[width=0.9\linewidth]{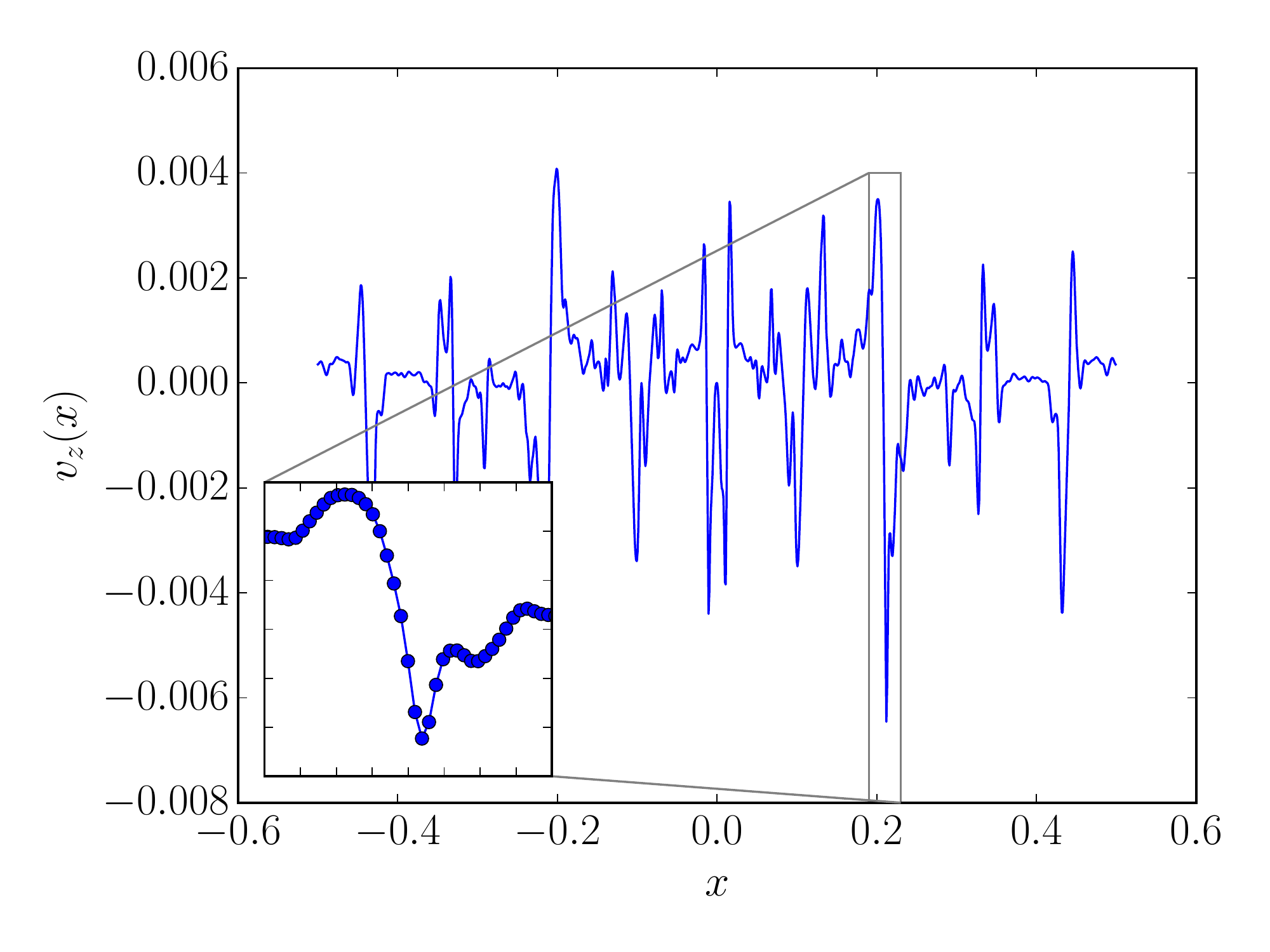}
	\vspace{-0.5cm}
    \caption{Vertical velocity as a function of $x$ measured at $z=0.1$ in run v7-1024-v at $t=600$. The inset zooms on the critical layer at $x\simeq 0.21$ with dots representing the spectral collocation points. Note the sharpness of the critical layers despite the large resolution.}
    \vspace{-0.2cm}
    \label{fig:vz}
\end{figure}

More interesting is the behaviour of these simulations when resolution
and dissipation processes are modified. To illustrate this, let us
consider higher resolution simulations with $1024^3$ Fourier modes. We
first perform a resolution test (h-1024) to reproduce our fiducial run
\emph{with hyper-diffusion} which confirms that our $256^3$ run is
numerically converged, at least with respect to $E_K$
(Fig.~\ref{fig:energy}, green line). We then restart this
high-resolution 
simulation at $t=400$ but revert to classical dissipation
coefficients. 
We consider two cases: $\mathrm{Re}=\mathrm{Pe}=10^6$
(v6-1024) and $\mathrm{Re}=\mathrm{Pe}=10^7$ (v7-1024). The
$\mathrm{Re}=10^6$ shows a clear and steep decay indicating that the
ZVI disappears for this Reynolds number. If we move up to
$\mathrm{Re}=10^7$, a decline is still seen, but we cannot say for
sure that the ZVI is deactivated. A careful examination of the critical
layers at $t=600$ in the $\mathrm{Re}=10^7$ case shows that they are
resolved by only 4 to 5 collocation points (Fig.~\ref{fig:vz}). We
therefore conclude that the critical Reynolds number $\mathrm{Re}_c$
 for the ZVI (if it exists) is certainly larger than $10^6$, and
possibly larger than $10^7$. Simulations with at least $2048^3$  
(or even $4096^3$) points will be required to confirm the existence of the ZVI with second-order dissipation operators. 

Such a high critical Reynolds number is actually expected from the phenomenology of subcritical transitions in shear flow \citep{L02}. In terms of thermodynamics, a shear flow is always trying to cancel the shear via various processes (viscous diffusion, turbulent transport, etc.). Phenomenologically, the flow switches from a laminar solution to an unstable solution when the ``turbulent'' diffusion it can get from the instability is larger than the viscous diffusion. This general argument is seen in Couette-Taylor experiments, subcritical transition in shear flows, etc. As shown by Longaretti (2002), this argument can be transformed into
\begin{align}
\mathrm{Re}_c\simeq \frac{S^2L^2}{\langle v_xv_y\rangle}	.
\end{align}
The measured turbulent transport in our fiducial run being $\langle v_xv_y\rangle\simeq 2\times 10^{-8}$, we expect $\mathrm{Re}_c\simeq 5\times 10^7$, surprisingly close to the limit found using brute force simulations. Although not definitely conclusive, all of these arguments point toward the fact that the ZVI requires tremendously high Reynolds numbers.

\subsection{Cooling and heating processes\label{sec:cool}}

The sensitivity to the Reynolds number indicates that the
ZVI mechanism is highly dependent on dissipation and diffusion. In protoplanetary
discs, Reynolds numbers are huge, so a high $Re_c$ is not physically a
problem (although it is definitely a problem for numerical
simulations). Cooling in these discs, on the other hand, is far from negligible. It is therefore desirable to test the existence of the ZVI in the presence of cooling and heating.

In protoplanetary discs, cooling and heating are dominated by radiative transfer, thermal conductivity being unimportant. If the length-scales $\lambda$ under consideration are larger than the photon mean free path $\ell_{\mathrm{ph}}=1/\kappa\rho$, where $\kappa$ is the opacity and $\rho$ is the gas density (i.e. the disc is optically thick on scale $\lambda$), cooling and heating can be approximated by thermal diffusion with a diffusion coefficient
\begin{align}
\label{eq:chi}
\chi=\frac{16\sigma T^3}{3\kappa\rho^2c_v},	
\end{align}
where $\sigma$ is the Stefan-Boltzmann constant and $c_v$ is the heat capacity of the gas, which we will assume to be diatomic. In the opposite optically thin limit $\lambda<\ell_{\mathrm{ph}}$, radiative cooling acts like a scale-free newtonian cooling with a characteristic timescale $t_c$
\begin{align}
\label{eq:cool}
t_c=\frac{\ell_\mathrm{ph}^2}{3\chi}	.
\end{align}
Note that \emph{this cooling timescale is not the same as the global cooling timescale} of a vertically integrated disc
(subject to external heating and radiative cooling), since here we look at \emph{small-scale thermal perturbations} embedded in an optically thick medium.

\begin{figure}
\vspace{-0.4cm}
	\includegraphics[width=0.9\linewidth]{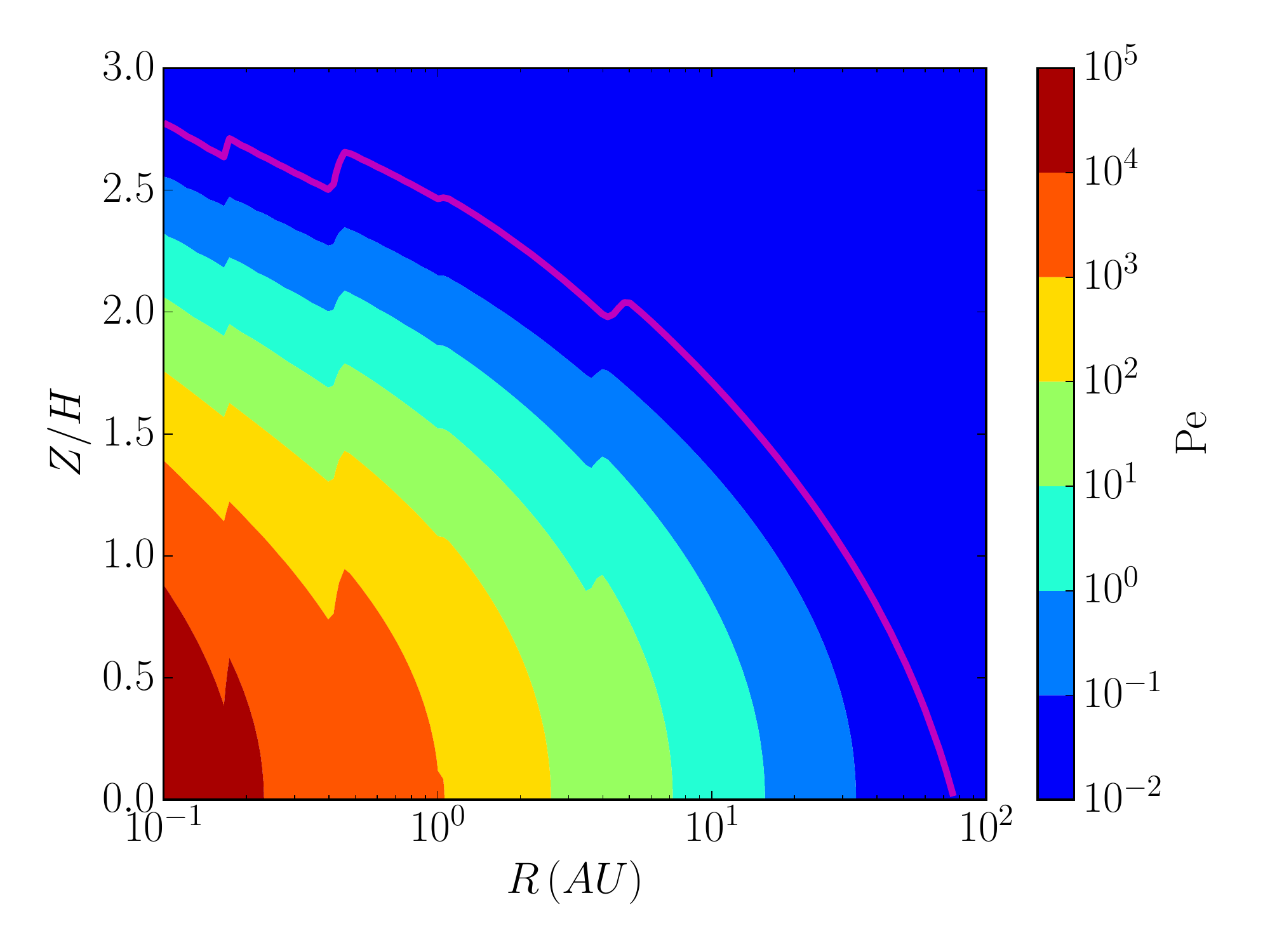}
	\vspace{-0.5cm}
    \caption{Peclet number as a function of position in a $0.01\,M_\odot$ disk model. The magenta contour defines the $\tau=1$ surface below which the disc is optically thick: $\ell_\mathrm{ph}<H$.}
    \label{fig:pe_disc}
\end{figure}

\begin{figure}
\vspace{-0.4cm}
	\includegraphics[width=0.9\linewidth]{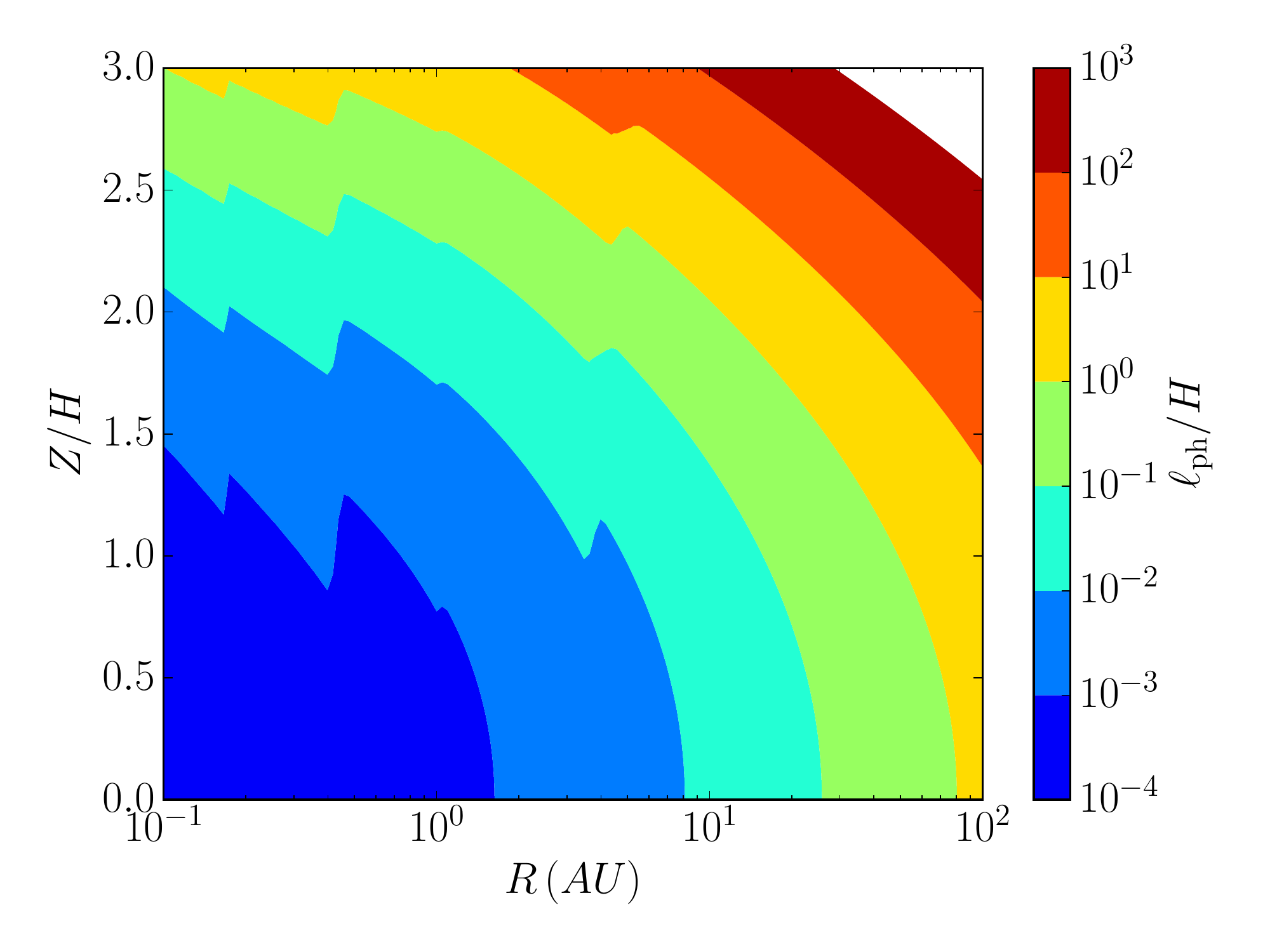}
	\vspace{-0.5cm}
    \caption{Photon mean free path $\ell_\mathrm{ph}$ compared to the disc scale height $H$ in a $0.01\,M_\odot$ disk model. Shortest mean free paths are found close to the midplane in the innermost parts of the disc.}
    \label{fig:mfp}
\end{figure}

\begin{figure}
\vspace{-0.4cm}
	\includegraphics[width=0.85\linewidth]{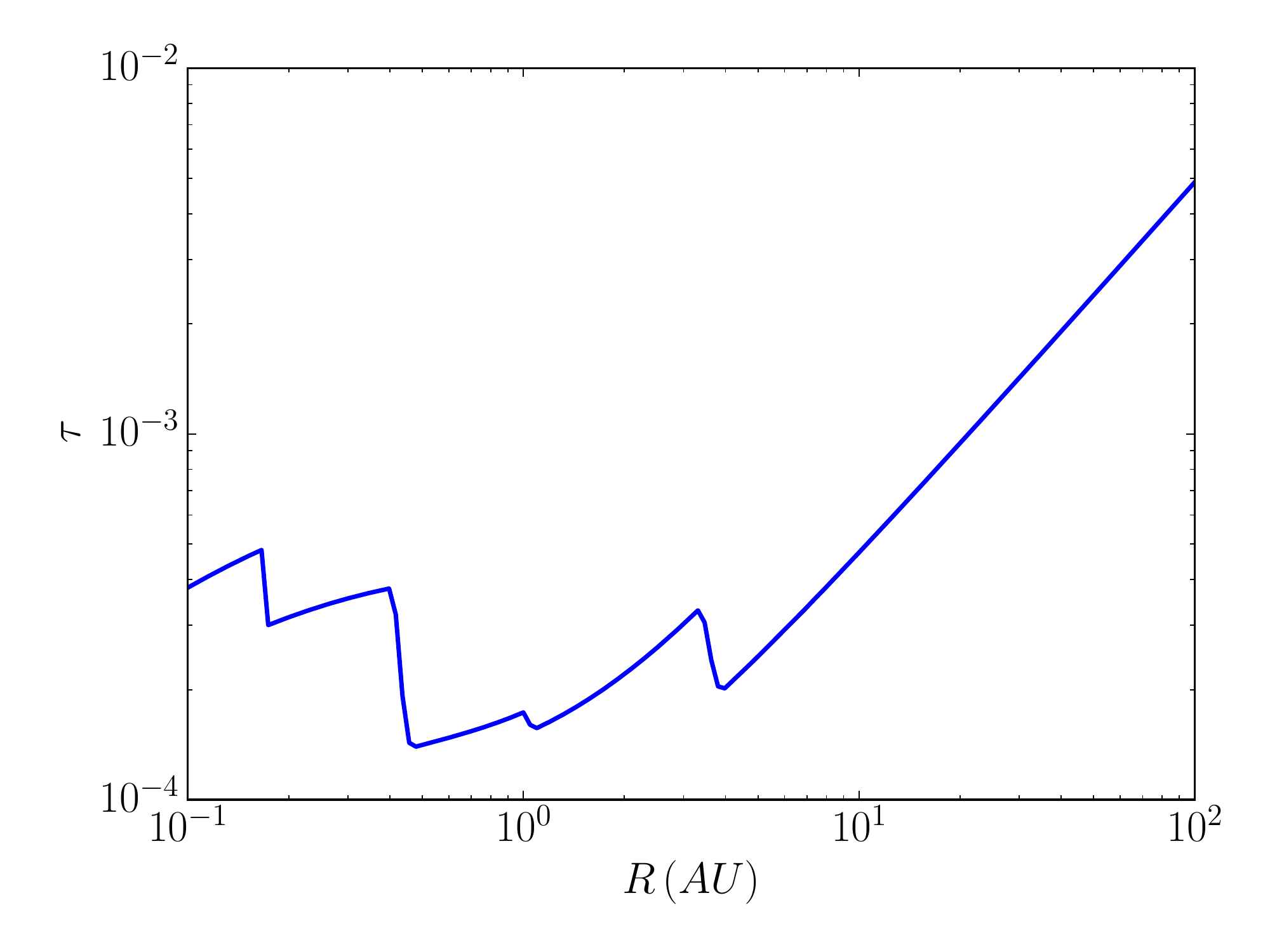}
	\vspace{-0.5cm}
    \caption{Dimensionless cooling time as a function of radius in a $0.01\,M_\odot$ disk model. Jumps are due to the condensation of molecules onto dust grains which abruptly change opacities \citep{S03}.}
    \label{fig:tau_disc}
    \vspace{-0.7cm}
\end{figure}

To illustrate the typical Peclet number and cooling times in
protoplanetary discs, we have considered a typical T-Tauri disc model
$\Sigma=140 R_\mathrm{AU}^{-1}\,\mathrm{g}\cdot\mathrm{cm}^{-2}$ and
$T=280\,R_\mathrm{AU}^{-1/2}\,\mathrm{K}$ which corresponds to a
$0.01\,M_\odot$ mass disc extending to $100\,\mathrm{AU}$. We assume
the disc to be vertically isothermal as we don't solve the full
radiative transfer equations. Rosseland opacities including gas and
dust contributions are obtained from \cite{S03} assuming spherical
homogeneous dust grains of solar composition. To compute the resulting
Peclet number, we have identified the box scale $L$ to the disc
pressure scale height\footnote{In principle $L$ can be
  arbitrarily smaller than $H$ since we work in the incompressible
  limit. We have not considered this case since it leads to critical
  disc $\mathrm{Pe}$ even larger than the one discussed here, leading
  to a smaller domain of existence for the ZVI.}  $H\equiv
c_s/\Omega$, where $c_s$ is the local sound speed and $\Omega$ the
local Keplerian frequency. The resulting map for thermal diffusion
(Peclet number) is shown in Fig.~\ref{fig:pe_disc}. As mentioned
above, the thermal diffusion approximation is valid only for scales
$\lambda>\ell_\mathrm{ph}$. The typical photon mean free path
$\ell_\mathrm{ph}$ is shown in Fig.~\ref{fig:mfp}. The smallest
$\ell_\mathrm{ph}/H$ is 
found close to the midplane in the inner parts of the disc. These are
the regions expected to be well described by the thermal diffusion
approximation on most relevant scales.\footnote{Note however that the thickness of the critical layers
involved in the ZVI can be several orders of magnitude smaller than the disc scale as it is set by gas molecular viscosity.
It is therefore possible that critical layers are always in the optically thin regime for realistic Reynolds numbers.}
On the contrary, the outer regions $R>10\,\mathrm{AU}$
have $\ell_\mathrm{ph}\lesssim H$. In these regions, cooling is best
described by a constant cooling time characterised by the
dimensionless parameter $\tau\equiv t_c\Omega$. Since the cooling time
(\ref{eq:cool}) does not depend on density, $\tau$ is only a function
of radius, shown in Fig.~\ref{fig:tau_disc}.  From these three
figures, 
we deduce that for $R\gtrsim 1\,AU$, $\mathrm{Pe}<10^3$ and $\tau<10^{-2}$.

\begin{figure}
	\includegraphics[width=0.9\linewidth]{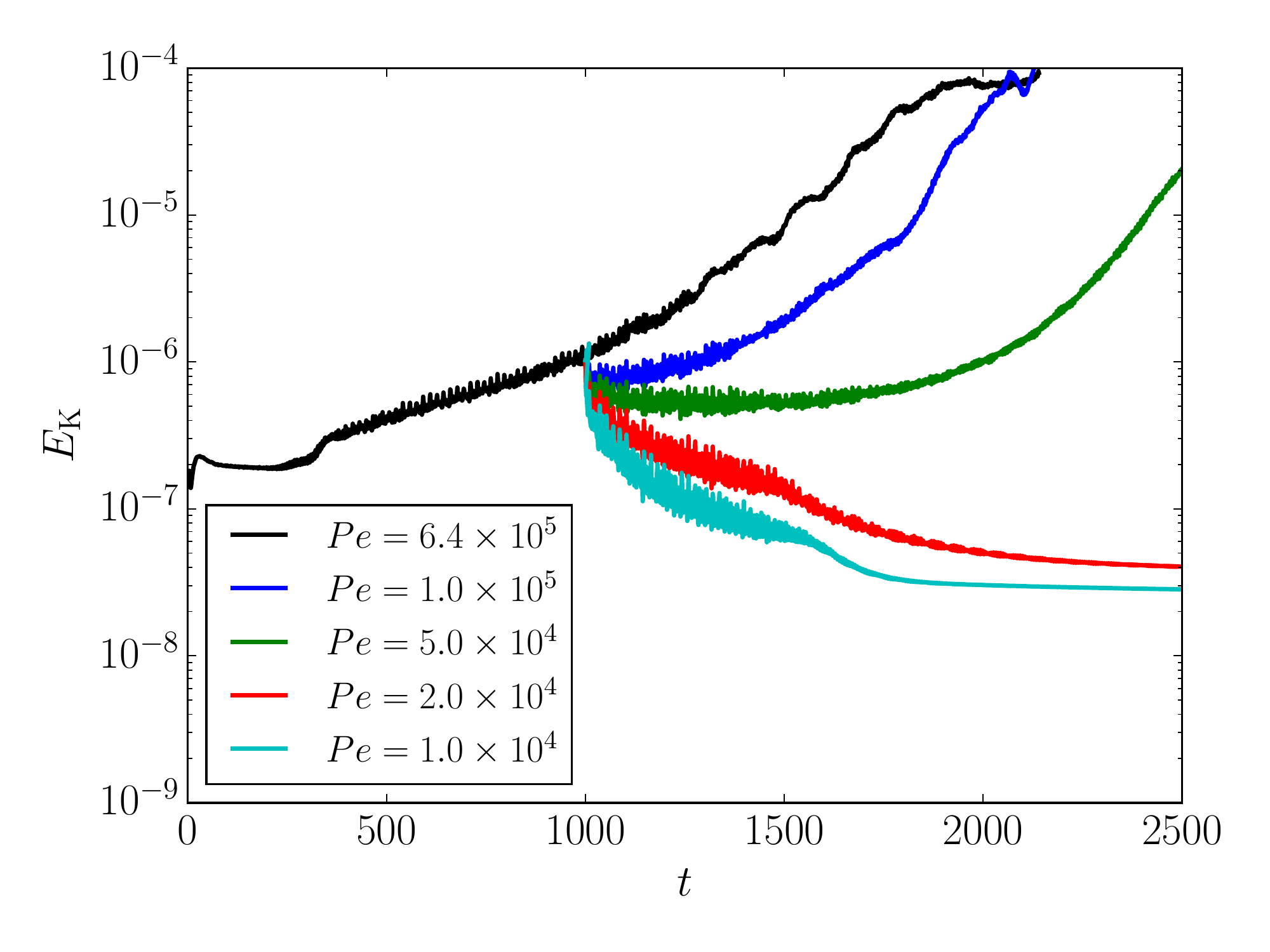}
	\vspace{-0.5cm}
    \caption{Volume averaged kinetic energy for runs dx-256-v varying thermal diffusivities. $\mathrm{Pe}>2\times 10^4$ is needed to sustain the ZVI. }
    \label{fig:pe_sim}
    \vspace{-0.4cm}
\end{figure}

\begin{figure}
	\includegraphics[width=0.9\linewidth]{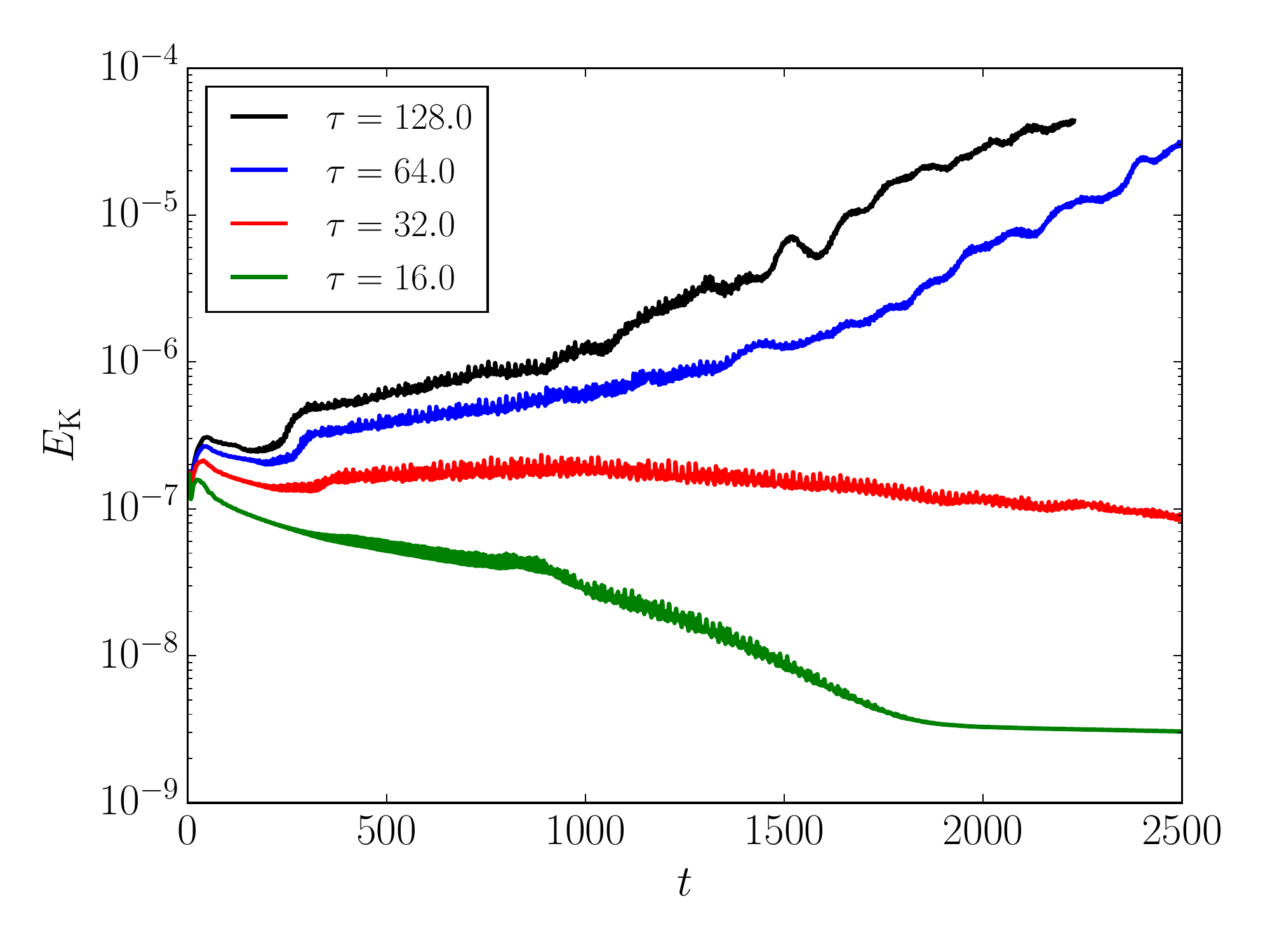}
	\vspace{-0.5cm}
    \caption{Volume averaged kinetic energy for runs tx-256-v varying thermal relaxation timescales. $\tau>16$ is needed to sustain the ZVI. }
    \label{fig:tau_sim}
\end{figure}

Our last task is to test in which parameter regime the ZVI lives. To
this end, we have performed a set of simulations identical to our
fiducial simulation, except that thermal hyper-diffusion is now
replaced by a classical thermal diffusion operator with
$\mathrm{Pe}\in [10^4,10^6]$ (runs dxxxx) or by a fixed cooling
parameter $\tau \in [10,200]$ (runs txxxx). We have used either the 
gaussian vortex initial conditions (runs ending with ``v'') or Kolmogorov
noise initial conditions (runs ending with ``k'').
The energy evolution of
the simulations starting with a gaussian vortex (Figs.~\ref{fig:pe_sim}-\ref{fig:tau_sim}) clearly
indicates that the ZVI requires $\mathrm{Pe}> 2\times 10^4$ and
$\tau>16$. Runs with $\tau\leq 16$ or $\mathrm{Pe}\leq10^4$ becomes
axisymmetric at $t\sim 2000$ which ensure that the ZVI is definitely  
switched off for this range of parameters. Very similar limits are obtained 
when using Kolmogorov noise as an initial condition (Tab.~\ref{tab:runs}). Our limits of existence for the ZVI 
therefore do not depend strongly on the chosen initial condition.

These dimensionless numbers are clearly excluded in our typical disc
model presented above, except maybe in the diffusive regime in the
innermost regions ($R\simeq 0.1 \mathrm{AU}$) which are also likely to
be unstable to the magneto-rotational instability due to their
proximity to the central star \citep{LB12}.

\section{Conclusions}

In this letter, we have explored the sensitivity of the zombie vortex
instability to diffusive and thermal processes. We find that one can
easily produce this instability with hyper-diffusion operators, but
not with classical viscous operators. We conjecture that a resolution
of at least $2048^3$ collocation spectral points and a Reynolds number
higher than $10^7$ are required to ascertain the presence of the ZVI
with physical dissipation. This should not come as a
surprise since the instability mechanism relies on the physics of buoyancy
critical layers, which are themselves controlled by diffusion (the process
that sets their characteristic lengthscale). 
It is therefore essential to properly resolve these structures with
realistic dissipation operators (i.e. neither hyper-diffusion nor
numerical dissipation). Note that finding the ZVI with  finite volume
codes does
 not solve this issue since these codes are also strongly affected by numerical diffusion.

We have also explored the sensitivity of the ZVI to cooling.
If radiative diffusion or Newtonian cooling is too efficient then the
action of buoyancy is diminished, as expected, and the instability
switches off. The critical Peclet number below which ZVI fails is
$\sim 10^4$, while the critical cooling time is $\sim 10\Omega^{-1}$.
This critical $\mathrm{Pe}$ have been obtained with a fixed
hyperdiffusivity so that the viscous scale is always much smaller than the thermal diffusion scale, as
in a real protoplanetary disc. However, the ZVI may also show a dependance on 
$\mathrm{Pr}=\nu/\chi$ or other combinations of dimensionless parameters. 
These dependancies have not been explored in this work.

Using a typical T-Tauri disc model of $0.01\,M_\odot$ mass, we find 
 that the ZVI may struggle to survive except in the
 densest and innermost regions of the disc ($R\sim 0.1\,\mathrm{AU}$)
 which are in any case MRI unstable. This is true whether the
 characteristic lengthscale of the ZVI falls in the diffusive or
 Newtonian cooling regimes. Taken on face value, these results 
cast doubt on the ZVI as a
 potential source of turbulent transport and vortices in late type
 objects (class II). We note, however, that younger discs ($M\sim 0.1
 M_\odot$) could reach $\mathrm{Pe}\sim 10^6$ at $R\sim 1\mathrm{AU}$
 thanks to the increase in gas density. These discs would be subject
 to gravitational instabilities in their outer part but could be ZVI
 unstable in their inner part. Nevertheless, this scenario must be
 confirmed by (a) demonstrating the existence and convergence of the
 ZVI with explicit viscous dissipation and (b) including a proper 
radiative transfer modelling to compute cooling accurately. 

Note finally that this work has been performed in a local approximation
(constant stratification, incompressibility, constant cooling). The ZVI being a local
instability \citep{MP13,MP15}, it is well captured and described by
this model.
Our study does not exclude the possibility of a \emph{global} instability which would be due
to the vertical structure of the disc. However, such a hypothetical instability would be driven 
by a different physical mechanism than that of the ZVI. Note also that
global simulations will inherit (and probably exacerbate) the ZVI's
numerical convergence problem (cf.\ Section 3.1). 

\section*{Acknowledgements}

The computations presented here were performed using the Froggy platform of the CIMENT infrastructure (https://ciment.ujf-grenoble.fr).
HNL acknowledges funding from STFC
grant ST/L000636/1 and helpful advice from John Papaloizou, Michael
McIntyre, and Steve Lubow.




\bibliographystyle{mnras}
\bibliography{zvi} 





\bsp	
\label{lastpage}
\end{document}